\documentclass[aps,prl,reprint,groupedaddress,preprintnumbers,nofootinbib,nobibnotes,amsmath,amssymb,nofootinbi]{revtex4-2}

\usepackage{amssymb,amsmath,amsfonts,amsbsy,graphicx,bm}
\usepackage{hyperref,xcolor,adjustbox}

\newcommand{\calO}{{\cal O}}

\newcommand{\pbh}{{\rm PBH}}
\newcommand{\fnl}{f_{\rm NL}}

\begin{document}

\preprint{APCTP-Pre2024-008}
\preprint{TU-1228}

\title{Non-Gaussianity from primordial black holes}

\author{Jinn-Ouk Gong$^{1,2}$}
\author{Naoya Kitajima$^{3,4}$}

\affiliation{
$^{1}$Department of Science Education, Ewha Womans University, Seoul 03760, Korea
\\
$^{2}$Asia Pacific Center for Theoretical Physics, Pohang 37673, Korea
\\
$^{3}$Frontier Research Institute for Interdisciplinary Sciences, Tohoku University, Sendai 980-8578, Japan
\\
$^{4}$Department of Physics, Tohoku University, Sendai 980-8578, Japan
}

\date{\today}

\begin{abstract}

We study the effects of non-Gaussianity from primordial black holes (PBHs). The formation of PBHs is in general a rare event and the number of PBHs fluctuates following the Poisson distribution function, which is independent from the pre-existing inflationary adiabatic fluctuations. Such fluctuations can dominate over the adiabatic mode on small scales. We focus on the non-Gaussianity of matter density fluctuations induced by the Poisson fluctuation of PBHs and discuss the potentially observable consequences such as the skewness, kurtosis and the scale-dependent bias.

\end{abstract}

\maketitle

\section{Introduction}

After the discovery of gravitational waves (GWs) from the binary black hole mergers by LIGO/Virgo Collaboration~\cite{LIGOScientific:2016aoc}, the possibility was suggested that those binary black holes are of primordial origin~\cite{Bird:2016dcv,Clesse:2016vqa,Sasaki:2016jop} -- primordial black holes (PBHs)~\cite{ZeldovichPBH,Hawking:1971ei,Carr:1974nx}. Typically, PBHs are formed when the amplitude of the primordial perturbations at certain length scale exceeds a threshold value so that as soon as such a region enters the horizon, gravitational collapse immediately follows and the corresponding horizon-size black hole is formed. Thus, the formation of PBHs accompanies rich phenomenologies in the early universe, such as induced GWs by the large amplitude of primordial perturbations \cite{Saito:2008jc,Bugaev:2010bb}. For implications of PBHs related to the GW astronomy, see e.g.~\cite{Sasaki:2018dmp}.

The threshold value of the primordial perturbations necessary for the formation of PBHs is $\mathcal{O}(10^{-1}-10^{-2})$~\cite{Harada:2013epa}, which is much higher than the value constrained on the cosmic microwave background scales, $\mathcal{O}(10^{-5})$~\cite{Planck:2018vyg}. Thus, we can notice the following two implications. First, the PBH formation is a rare event and accordingly PBHs are sparsely distributed in space. So, the distribution of PBH follows that for random, independent events -- the Poisson distribution. Thus, there is a Poisson noise in the number of PBHs~\cite{Meszaros:1975ef}. It causes an additional fluctuation of the density perturbation as an isocurvature mode~\cite{Afshordi:2003zb}, which we will call ``PBH Poisson fluctuation''. It can enhance the structure formation on small scales at high redshift~\cite{Inman:2019wvr} and hence the PBH abundance can be constrained by small-scale observations such as the Ly$\alpha$ forest~\cite{Afshordi:2003zb,Carr:2018rid} and 21cm forest~\cite{Gong:2017sie,Gong:2018sos,Mena:2019nhm}.

Second, the large amplitude necessary for the PBH formation means that the PBH abundance is sensitively dependent on the tail of the distribution which the primordial perturbations follow. Since the statistics of the primordial perturbations is consistent with Gaussian distribution~\cite{Planck:2019kim}, any small deviation from Gaussian statistics, viz. non-Gaussianity can significantly affect the PBH abundance by slightly changing the shape of the distribution tail~\cite{Bullock:1996at}. This is why the impacts of the primordial non-Gaussianity on PBHs have been extensively studied~\cite{Byrnes:2012yx,Young:2013oia,Young:2014oea,Tada:2015noa,Franciolini:2018vbk,Kitajima:2021fpq}.

On the other hand, as discussed above, the PBH number fluctuation itself follows the Poisson distribution that highly deviates from the Gaussian one, even if the initial fluctuations responsible for the PBH formation are perfectly Gaussian. This implies non-Gaussian features are inevitably predicted in any PBH formation scenario. Thus, we need to include correctly these non-Gaussian effects on small-scale structure formation when the PBH Poisson fluctuation becomes important. We particularly focus on the three- and four-point correlation functions of density fluctuation, related to the bispectrum and trispectrum respectively, as the leading and next-to-leading order corrections to the Gaussian case and calculate the skewness and kurtosis as potentially observable quantities. We also show the scale-dependent bias shift induced by the PBH Poisson fluctuations.

\section{Non-Gaussianity from PBHs}

\subsection{Poisson fluctuation of the number of PBHs}

The Poisson noise of the number of PBHs gives an additional contribution to the density perturbation independent of the adiabatic fluctuation originated during inflation. The density contrast is then given by the following summation:
\begin{equation}
\delta(\bm{x}) = \delta_{\rm adi}(\bm{x})+\delta_P(\bm{x}) \, ,
\end{equation}
where $\delta_{\rm adi}(\bm{x})$ and $\delta_P(\bm{x})$ corresponds respectively to the adiabatic mode and the PBH Poisson fluctuation. We take these two variables as random fields with no correlation to each other.

The Poisson fluctuation $\delta_P(\bm{x})$ is originated from the fluctuation of the number of PBHs, $N_\pbh$, in each comoving patch of the Universe, which follows the Poisson distribution function:
\begin{equation}
P(N_\pbh) = \frac{\lambda^{N_\pbh}e^\lambda}{N_\pbh!} \, ,
\end{equation}
where $\lambda \equiv \bar{N}_\pbh$ is the mean and also the variance of $N_\pbh$. Denoting the fluctuation of $N_\pbh$ as $\delta N_\pbh = N_\pbh-\bar{N}_\pbh$, the statistical properties of the Poisson distribution read
\begin{align}
\label{eq:P-property}
\big\langle \delta N_\pbh^2 \big\rangle = \big\langle \delta N_\pbh^3 \big\rangle = \big\langle \delta N_\pbh^4 \big\rangle_c
= \bar{N}_\pbh \, ,
\end{align}
where the angular brackets denote the statistical average and $\langle X^4 \rangle_c \equiv \langle X^4 \rangle-3 \langle X^2 \rangle^2$. This fluctuation generates an additional CDM isocurvature mode: 
\begin{equation}
\label{eq:iso}
S = f_\pbh\frac{\delta{N}_\pbh}{\bar{N}_\pbh} \, ,
\end{equation}
with $f_\pbh \equiv \Omega_{\rm PBH}/\Omega_{\rm DM}$ being the fraction of PBHs in total dark matter.  From \eqref{eq:P-property}, we obtain
\begin{align}
\langle S^2 \rangle = \frac{f_\pbh^2}{\bar{N}_\pbh} \, , \quad
\langle S^3 \rangle = \frac{f_\pbh^3}{\bar{N}_\pbh^2} \, , \quad
\langle S^4 \rangle_c = \frac{f_\pbh^4}{\bar{N}_\pbh^3} \, .
\end{align}
Note that $S(\bm{x})$ corresponds to an initial value of $\delta_P(\bm{x})$.

Since $N_\pbh$  in each region can be regarded as an independent random variable, we can assume that $S(\pmb{x})$ has no spatial correlation and thus the spatial $n$-point correlation function takes the following expression:
\begin{align}
\left\langle \prod_i^n S(\bm{x}_i) \right\rangle_c 
= V^{n-1} \prod_{i=1}^{n-1} \delta^{(3)}(\bm{x}_i-\bm{x}_{i+1}) \langle S^n \rangle_c\,,
\end{align}
which gives the following correlation function in Fourier space:
\begin{align}
\left\langle \prod_i^n S(\bm{k}_i) \right\rangle_c 
= (2\pi)^3 \delta^{(3)} \left( \sum_i^n \bm{k}_i \right) V^{n-1} \langle S^n \rangle_c\,.
\end{align}
From above, one obtains the following scale-invariant power spectrum from two-point correlation function:
\begin{equation} \label{eq:PS}
P_S = V\langle S^2 \rangle = \frac{f_\pbh^2}{\bar{n}_\pbh} 
\approx 2.9 \times 10^{-11} {\rm Mpc}^3 f_\pbh \bigg( \frac{M_\pbh}{M_\odot} \bigg) \, ,
\end{equation}
where 
\begin{equation}
\bar{n}_\pbh 
= \frac{f_\pbh\Omega_{\rm DM}\rho_{c0}}{M_\pbh} 
\approx 3.3 \times 10^{10} {\rm Mpc}^{-3} f_\pbh \frac{M_\odot}{M_\pbh}
\end{equation} 
is the comoving number density of PBHs \cite{Afshordi:2003zb,Gong:2017sie}. Similarly, the bispectrum and trispectrum can be obtained as follows:
\begin{align}
\label{eq:biS}
B_S & = \frac{f_\pbh^3}{\bar{n}_\pbh^2}
\approx 8.9 \times 10^{-22} {\rm Mpc}^6 f_\pbh \bigg( \frac{M_\pbh}{M_\odot} \bigg)^2
\, ,
\\
T_S & = \frac{f_{\rm PBH}^4}{\bar{n}_{\rm PBH}^3} \approx 2.6\times 10^{-32}{\rm Mpc}^9 f_{\rm PBH}  \left(\frac{M_{\rm PBH}}{M_\odot}\right)^3
\, .
\end{align}

\subsection{Skewness and kurtosis}

The effect of non-Gaussianity is imprinted to the higher cumulants of the density contrast beyond the variance. Specifically, the bispectrum and trispectrum give third and fourth cumulants respectively. The third cumulant of the density contrast smoothed over the scale $R$ can be calculated as
\begin{align}
\begin{split}
\langle \delta_R^3 \rangle_c 
& = 
\frac{B_S}{8\pi^4} \int^\infty_0 dk_1 k_1^2 \mathcal{M}_R(k_1)  \int^\infty_0 dk_2 k_2^2 \mathcal{M}_R(k_2) 
\\
& \quad 
\times \int^1_{-1}d\mu \mathcal{M}_R(|\bm{k_1}-\bm{k_2}|)
\, ,
\end{split}
\end{align}
with $\mu \equiv \bm{k}_1\cdot\bm{k}_2/(k_1 k_2)$ and $\mathcal{M}_R(k)$ is defined through $\delta_R(\bm{k}) = \mathcal{M}_R(k) S(\bm{k})$ which includes the transfer function, linear growth factor and window function.
The fourth cumulant can be calculated as well:
\begin{align}
\begin{split}
\langle \delta_R^4 \rangle_c 
&= 
T_S \int \frac{d^3k_1}{(2\pi)^3} \int \frac{d^3k_2}{(2\pi)^3} \int \frac{d^3k_3}{(2\pi)^3} 
\\
&\quad 
\times {\cal M}_R(k_1){\cal M}_R(k_2){\cal M}_R(k_3) {\cal M}_R(|\bm{k}_1+\bm{k}_2+\bm{k}_3|)
\, .
\end{split}
\end{align}
Note that for small-scale limit (typically $R < 1 \, {\rm Mpc}$), the above integral can be analytically calculated as the transfer function for the CDM isocurvature mode is constant~\cite{Afshordi:2003zb}.

\begin{figure}[tp]
\centering
\includegraphics [width = 7.5cm, clip]{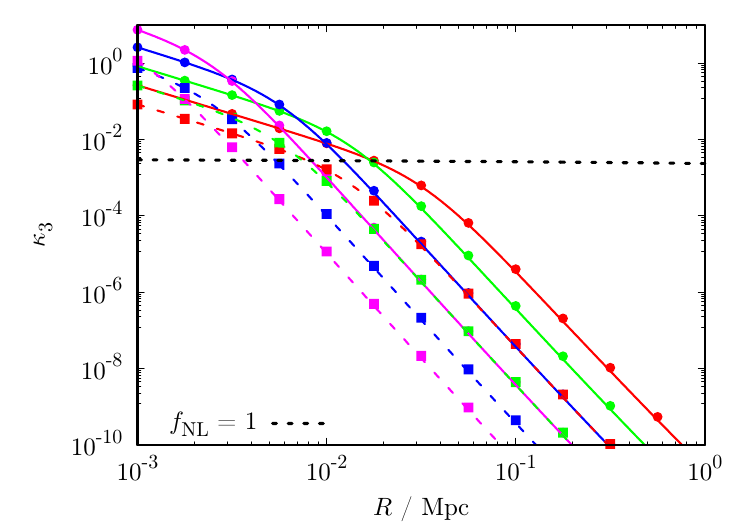}
\label{subfig:kappa3}
\includegraphics [width = 7.5cm, clip]{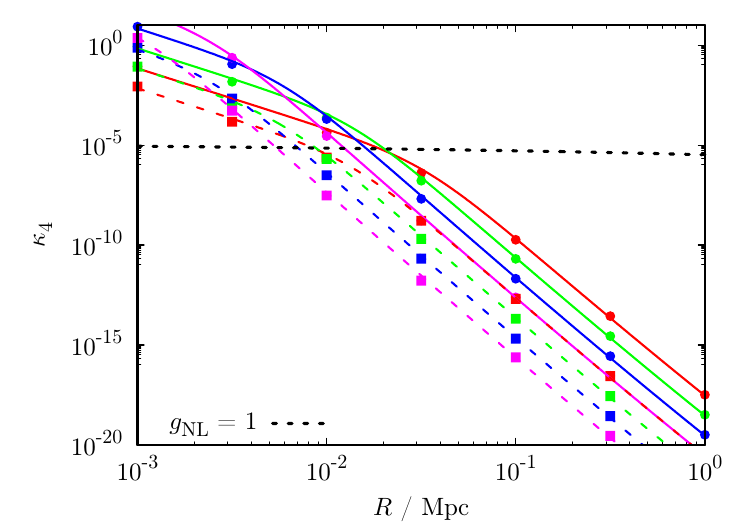}
\label{subfig:kappa4}
\caption{
The skewness $\kappa_3$ (top) and the kurtosis $\kappa_4$ (bottom) of the density contrast smoothed over the scale $R$. Points and lines correspond respectively to the numerical result and the analytic approximation. We have taken $M_{\rm PBH}=1M_\odot$ (dashed), $10M_\odot$ (solid) and $f_{\rm PBH}=1$ (red), $0.1$ (green), $0.01$ (blue), $0.001$ (magenta). The dotted black lines show the fitting formulas with $\fnl=1$ (top) and $g_{\rm NL} =1$ (bottom).
}
\label{fig:kappa}
\end{figure}

The reduced cumulant\footnote{
Another definition of the reduced cumulant is $S_n(R) \equiv \langle \delta^n_R \rangle_c / \langle\delta^2_R\rangle^{n-1}$.} 
of smoothed density contrast is defined by $\kappa_n(R) \equiv \langle \delta^n_R \rangle_c/\sigma_R^n$ with $\sigma_R^2 = \langle \delta_R^2 \rangle^{1/2}$. Figure~\ref{fig:kappa} shows the reduced skewness $\kappa_3$ (top) and the kurtosis $\kappa_4$ (bottom) of the smoothed density contrast. Using the fitting formula, $\kappa_3 \simeq 3.14\times 10^{-3}\fnl \sigma_R^{0.165}$ and $\kappa_4 \simeq 1.14\times 10^{-5}g_{\rm NL} \sigma_R^{0.73}$  \cite{Chongchitnan:2010xz}, we also show the case for the primordial local-type non-Gaussianity with $\fnl=1$ and $g_{\rm NL} = 1$. It shows that the non-Gaussianity from PBH becomes significant for the scale $R \lesssim 0.01\,{\rm Mpc}$.

\subsection{Effective non-linear parameters}

The isocurvature perturbation \eqref{eq:iso} can be written as the difference between the curvature perturbation in dark matter and that in radiation:
\begin{equation}
S = 3 (\zeta_m - \zeta_r) \, ,
\end{equation}
where $\zeta_m$ ($\zeta_r$) is the curvature perturbation on the uniform matter (radiation) hypersurfaces. Without PBHs, both dark matter and radiation are of the same origin (e.g. from inflation) and $\zeta_m = \zeta_r$. Now, during matter domination, the total curvature perturbation $\zeta_m=\zeta$ can be written as
\begin{equation}
\zeta = \zeta_\text{inf} + \frac{1}{3}S \, ,
\end{equation}
where we have replaced $\zeta_r$ with $\zeta_\text{inf}$, which is the primordial curvature perturbation generated during inflation. Assuming that there is no correlation between $\zeta_{\rm inf}$ and $S$, the power spectrum of curvature perturbation is given by
\begin{equation} 
\label{eq:Pzeta}
P_\zeta (k) = P_\text{inf}(k) + \frac{1}{9}P_S 
= \left\{
\begin{array}{ll}
\dfrac{1}{9}P_S & \text{for } k>k_\star 
\\
\dfrac{1}{9}P_S \bigg( \dfrac{k}{k_\star} \bigg)^{-3} & \text{for } k<k_\star
\end{array}
\right. ,
\end{equation}
where $P_{\rm inf}(k)$ is the power spectrum of $\zeta_{\rm inf}$ and $k_\star$ is defined by $P_\text{inf}(k_\star) = P_S/9$. Note that $P_S$ is a constant value given by \eqref{eq:PS}.

Let us assume that $\zeta_\text{inf}$ is perfectly Gaussian and only $S$ is responsible for non-Gaussianity. Then, the bispectrum of the total curvature perturbation is
\begin{equation}
B_\zeta(k_1,k_2,k_3) = \frac{1}{27}B_S \, ,
\end{equation}
where $B_S$ is a constant given by \eqref{eq:biS}. On the other hand, the bispectrum of curvature perturbation can be characterized by the non-linear parameter $\fnl$ as 
\begin{equation}
B_\zeta(k_1,k_2,k_3) = \frac{6}{5}\fnl \big[ P_\zeta(k_1)P_\zeta(k_2) + \text{2 perms} \big] \, .
\end{equation}
Specifically, for $k \equiv k_3 \ll k_1 \approx k_2$ and $k_1$, $k_2 \gg k_\star$, we obtains the following ``effective'' scale-dependent $\fnl$:
\begin{equation}
\fnl \approx 
\begin{cases}
\cfrac{5}{2f_\pbh} \left[ 1 + 2 \bigg( \cfrac{k}{k_\star} \bigg)^{-3} \right]^{-1} ~&\text{for}~~k < k_\star 
\\[3mm]
\cfrac{5}{6f_\pbh} ~&\text{for}~~k>k_\star
\end{cases}
\, .
\end{equation}
Note that the effective $\fnl$ can be significantly large for $f_\pbh \ll 1$ on small scales.

Now let us consider the trispectrum of curvature perturbation, $T_\zeta = T_S/81$.
One can also express $T_\zeta$ with two non-linear parameters $g_{\rm NL}$ and $\tau_{\rm NL}$ as follows:
\begin{widetext}
\begin{align}
T_\zeta(k_1,k_2,k_3,k_4) =
\frac{54}{25}g_{\rm NL} \big[ P_\zeta(k_1)P_\zeta(k_2)P_\zeta(k_3) + \text{3 perms} \big]
+ \tau_{\rm NL} \Big\{ P_\zeta(k_1)P_\zeta(k_2) \big[ P_\zeta(k_{13}) + P_\zeta(k_{14}) \big] + \text{11 perms} \Big\},
\end{align}
\end{widetext}
where $k_{ij} \equiv |\bm{k}_i+\bm{k}_j|$. Here, $g_{\rm NL}$ and $\tau_{\rm NL}$ are also scale-dependent as $\fnl$. To see this explicitly, let us first consider the configuration $k=k_4 \ll k_1 \approx k_2 \approx k_3$ and $k_1$, $k_2$, $k_3 \gg k_\star$. Then, we find
\begin{equation}
\label{eq:Tzeta}
T_\zeta = 
\begin{cases}
\dfrac{P_S^3}{9^3} \Bigg[ \dfrac{162}{25} g_{\rm NL} \bigg( \dfrac{k_\star}{k} \bigg)^{3}
+ 12\tau_{\rm NL} \bigg( \dfrac{k_\star}{k} \bigg)^{3} \Bigg]
~&\text{for}~~k \ll k_\star 
\\[3mm]
\dfrac{P_S^3}{9^3} \bigg( \dfrac{216}{25} g_{\rm NL} +24\tau_{\rm NL} \bigg)
~&\text{for}~~k \gg k_\star
\end{cases}
\, .
\end{equation}
and correspondingly
\begin{alignat}{3}
g_{\rm NL}+\cfrac{50}{27}\tau_{\rm NL} & \simeq \cfrac{25}{18 f_\pbh^2} \bigg(\cfrac{k}{k_\star}\bigg)^3 && \quad \text{for}~~k \ll k_\star \, ,
\\
g_{\rm NL}+\cfrac{25}{9}\tau_{\rm NL} & \simeq \cfrac{25}{24 f_\pbh^2} && \quad \text{for}~~k \gg k_\star
\, ,
\end{alignat}
which implies that $g_{\rm NL}$ and/or $\tau_{\rm NL}$ can be significantly larger than $\fnl$ for $f_\pbh \ll 1$. 

Let us consider another configuration $k = k_{12} \approx k_{34} \ll k_1 \approx k_2 \approx k_3 \approx k_4$.
In this case, we obtain
\begin{align} 
\label{eq:Tzeta-b}
T_\zeta = \frac{P_S^3}{9^3} \Bigg\{ \frac{216}{25} g_{\rm NL}
+ 8\tau_{\rm NL} \left[ 2 + \bigg( \frac{k_\star}{k} \bigg)^{3} \right] \Bigg\},
\end{align}
for $k \ll k_\star$ and thus
\begin{alignat}{3}
\tau_{\rm NL} & \simeq \cfrac{9}{8 f_\pbh^2} \bigg(\cfrac{k}{k_\star}\bigg)^3 && \quad \text{for}~~k \ll k_\star \, , 
\\
g_{\rm NL}+\cfrac{25}{9}\tau_{\rm NL} & \simeq \cfrac{25}{24 f_\pbh^2} && \quad \text{for}~~k \gg k_\star \, .
 \end{alignat}
Note that the $\tau_{\rm NL}$ term dominates for $k \ll k_\star$ and we have neglect the $g_{\rm NL}$ term in the right-hand side in \eqref{eq:Tzeta-b}, provided that $g_{\rm NL}$ is not too large.

\subsection{scale-dependent bias}

It is well-known that non-Gaussianity can also lead to a scale-dependent bias shift to the halo power spectrum \cite{Dalal:2007cu,Matarrese:2008nc,Slosar:2008hx,Chongchitnan:2010xz,Gong:2011gx}. Expanding the real-space two-point correlation function of haloes $\xi_h(\bm{x}_1,\bm{x}_2)$~\cite{Politzer:1984nu,Grinstein:1986en,Matarrese:1986et}, the leading and next-to-leading order non-Gaussian corrections\footnote{
We drop the term $\big[\xi^{(2)}_R(\bm{x}_1,\bm{x}_2)\big]^2/2$ in the right-hand side of \eqref{eq:Delta_xih4}. Note that this term exists even without the intrinsic non-Gaussian correction (due to the bispectrum and the trispectrum) and causes some problems in the calculation of the halo power spectrum which necessitate fully non-linear analysis \cite{Gong:2011gx}. This is beyond the scope of this work and here we simply neglect it by assuming that this term can be separable from the intrinsic non-Gaussian contribution.} 
are given respectively by 
\begin{align} 
\label{eq:Delta_xih3}
\Delta\xi_h^{(3)}(\bm{x}_1,\bm{x}_2) 
& = \left(\frac{\nu_R}{\sigma_R}\right)^3 \xi_R^{(3)}(\bm{x}_1,\bm{x}_1,\bm{x}_2) \, ,
\\
\label{eq:Delta_xih4}
\Delta\xi_h^{(4)}(\bm{x}_1,\bm{x}_2) 
& = \left(\frac{\nu_R}{\sigma_R}\right)^4 \bigg[ \frac{1}{3}\xi_R^{(4)}(\bm{x}_1,\bm{x}_1,\bm{x}_1,\bm{x}_2) 
\nonumber\\
& \hspace{5em} 
+\frac{1}{4}\xi_R^{(4)}(\bm{x}_1,\bm{x}_1,\bm{x}_2,\bm{x}_2) \bigg] \, ,
\end{align}
where $\nu_R = \delta_c/\sigma_R$ and $\xi^{(n)}_R$ is the $n$-point correlation function.

Then, the non-Gaussian correction to the halo power spectrum can be calculated as
\begin{align}
\Delta P_h^{(n)} = \int d^3x e^{-\bm{k}\cdot\bm{x}} \Delta \xi^{(n)}_h(|\bm{x}|) \, .
\end{align}
The halo (Eulerian) bias is given by $b_h = 1+b_L$, with $b_L$ the Lagrangian bias defined by $b_L^2 = P_h(k,z)/P_\delta(k,z)$. Without non-Gaussian corrections, it is simply given by $b_L = \delta_c/(\sigma_R^2(z))$. Thus we obtain the scale-dependent non-Gaussian correction to the halo bias:
\begin{align}
\frac{\Delta b_h^{(n)}}{b_h} = \frac{1}{2b_h (b_h-1)} \frac{\Delta P_h^{(n)}(k,z)}{P_\delta(k,z)} \, ,
\end{align}
where $b_h$ in the above expression is the Gaussian linear bias. We have directly calculated the non-Gaussian shift of the halo power spectrum up to the next-to-leading order contributions by numerically integrating the correlation functions \eqref{eq:Delta_xih3} and \eqref{eq:Delta_xih4}.

\begin{figure}[tp]
\centering
\includegraphics [width = 7.5cm, clip]{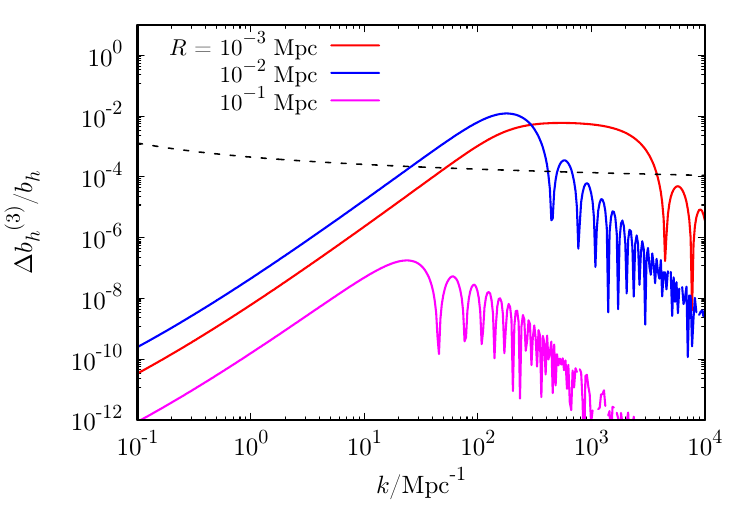}
\label{subfig:Deltabh3}
\includegraphics [width = 7.5cm, clip]{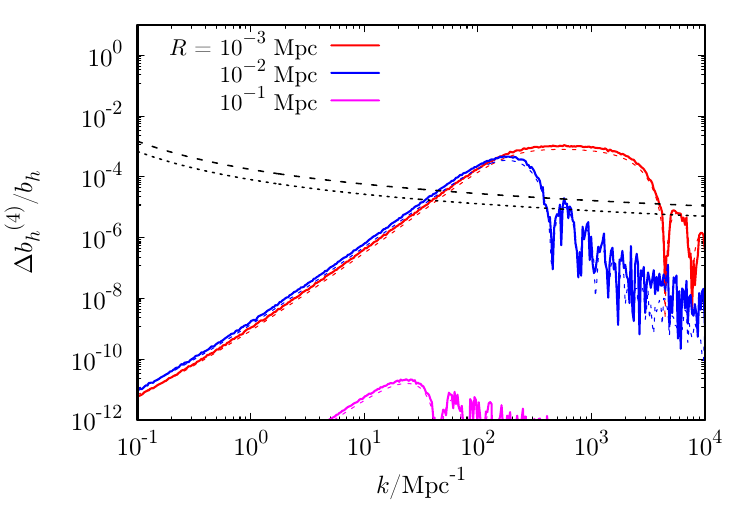}
\label{subfig:Deltabh4_1d}
\caption{
The leading and next-to-leading order non-Gaussian corrections to the the halo bias, $\Delta b_h^{(3)}(k)/b_h$ (top) and $\Delta b_h^{(4)}(k)/b_h$ (bottom) with smoothing scale $R=10^{-3}$ Mpc (red), $10^{-2}$ Mpc (blue) and $10^{-1}$ Mpc (magenta) for $z=20$, $M_{\rm PBH}=10M_\odot$ and $f_\pbh = 10^{-2}$.
In the bottom panel, the solid and dotted lines correspond to the first and second terms in the square bracket on the right-hand side of \eqref{eq:Delta_xih4}, but these two lines almost coincide.
For comparison, we also show the predictions from the primordial local-type non-Gaussianity with (top) $\fnl=1$ (dashed) and (bottom) $g_{\rm NL}=1000$ (black-dashed) and $\tau_{\rm NL}=1000$ (black-dotted).
}
\label{fig:Deltabh}
\end{figure}

Figure~\ref{fig:Deltabh} shows the results together with the predictions from primordial local-type non-Gaussianity with constant non-linear parameters $\fnl$~\cite{Dalal:2007cu,Matarrese:2008nc,Slosar:2008hx}, $g_{\rm NL}$ and $\tau_{\rm NL}$~\cite{Gong:2011gx}. The non-Gaussian shift induced by the PBH Poisson fluctuation can be significantly larger than that by the primordial local-type case with $\fnl=1$ for small scales, typically $R \lesssim 10^{-2}$ Mpc. Further, the bias corrections exhibits a peak, diminishing on large scales because of the scale-dependent non-Gaussianity, and oscillating on small scales because of the transfer function. Note that the perturbative expansion remains valid in this case, i.e. $\Delta b_h^{(3)} \gg \Delta b_h^{(4)}$ is satisfied. However, for smaller scales, these two contributions become comparable and the perturbative expansion may not be applied (see also Figure~\ref{fig:kappa}). Thus, we implicitly assume the smoothing scale $R \gtrsim 10^{-3}$ Mpc for our analysis to work.

\begin{figure}[tp]
\centering
\includegraphics [width = 7.5cm, clip]{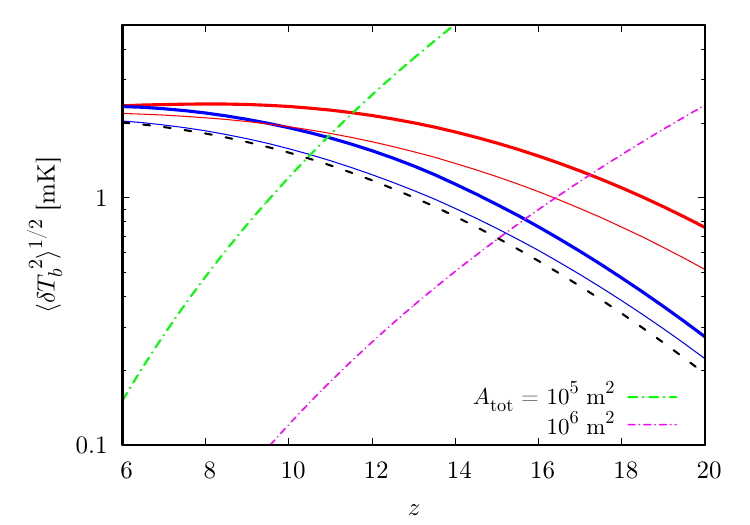}
\caption{
The root-mean-square of the 21cm brightness temperature.
The thick (thin) solid lines correspond to the predicted signals with (without) scale-dependent bias shift due to the PBH-induced non-Gaussianity for $f_\pbh = 10^{-3}$ (red), $10^{-4}$ (blue) and $M_\pbh = 30M_\odot$. The dashed black line represents the case without PBH (Poisson fluctuations) and the two dash-dotted  lines show the expected sensitivity by SKA-like observations with a typical observational setup~\cite{Gong:2017sie} parametrized by the total effective area parameter $A_{\rm tot}$.
}
\label{fig:T_21cm_rms}
\end{figure}

As a specific example in which the scale-dependent bias derived above takes a significant role, we consider the 21cm emission signature from minihalos at high redshift~\cite{Iliev:2002ms}. The root-mean-square of the 21cm brightness temperature fluctuation $\langle\delta T_b^2\rangle^{1/2}$ can be a detectable signature for radio telescope observations such as the Square Kilometer Array (SKA) and the Fast Fourier Transform Telescope (FFTT). In particular, non-Gaussianity affects the emission signals through the flux-averaged bias \cite{Chongchitnan:2012we,Sekiguchi:2018kqe}. Figure~\ref{fig:T_21cm_rms} shows the modification of $\langle\delta T_b^2\rangle^{1/2}$ by the PBH-induced non-Gaussianity. We can clearly see that the change is almost $\calO(1)$ mK, well within observational sensitivity, and thus it can be a unique signature of the PBH-induced non-Gaussianity.

\section{Conclusions and discussions}

In this article, we have reported that the scale-dependent non-Gaussianity of matter density fluctuation is inevitably induced from the Poisson fluctuation of PBHs. The effect can be described by the scale-dependent effective non-linear parameters such as $\fnl$, $g_{\rm NL}$ and $\tau_{\rm NL}$. On small scales, typically smaller than $10^{-2}$ Mpc, those non-linear parameters can be larger than $\mathcal{O}(1)$ with PBH mass $M_\pbh = 1$--$100M_\odot$.

Thus, the detection of such strongly scale-dependent non-Gaussianity can be a unique signature of the PBH formation scenario. For instance, scale-dependent bias corrections are significant and peaked on small scales, typically around 10 kpc scales and it would lead to the significant modification of the number of small dark matter haloes on those scales, with typical mass $10^3-10^7 M_\odot$ at high redshift~\cite{Gong:2017sie}.

Small-scale matter density fluctuations can be probed by the observation of the dark matter subhaloes in the Milky Way galaxy. For instance, dark matter subhaloes passing through the galaxy, with the typical mass $10^7 M_\odot$, affect the phase space distribution of stars~\cite{Buschmann:2017ams}. Astrometry by means of the weak gravitational lensing can also a powerful probe of the population of dark matter subhaloes~\cite{Mishra-Sharma:2020ynk}. The subhalo kinematics can also leave a signal in pulsar timing observations and the small-scale primordial fluctuations for $k >10^3~{\rm Mpc}^{-1}$ can be constrained~\cite{Lee:2020wfn}. In addition, the observations of the number of dwarf spheroidal galaxies and the stellar stream can be promising probes of dark matter substructure and the primordial density fluctuation on scales $k\sim 10^2$--$10^3~{\rm Mpc}^{-1}$ can be constrained~\cite{Ando:2022tpj}.

The characteristic sub(mini)halo abundance is imprinted as a signature of the PBH formation scenario and can potentially be searched by the  observations mentioned above. The non-Gaussian feature discussed in this work is a unique prediction, which makes it distinguishable from other possible scenarios, e.g. a simple enhancement of the inflationary small-scale fluctuations. In particular, the constraint on the PBH abundance from the subhalo observations needs a careful calculation of the halo mass function with the inclusion of the scale-dependent bias. We leave it for future studies.

\section{acknowledgments}

%
This work is supported in part by the National Research Foundation grants 2019R1A2C2085023 (JG) and the Japan Society for the Promotion of Science KAKENHI Grant Numbers 20H01894 (NK), 20H05851 (NK), 21H01078 (NK), 21KK0050 (NK).
JG also acknowledges the Korea-Japan Basic Scientific Cooperation Program supported by the National Research Foundation of Korea and the Japan Society for the Promotion of Science (2020K2A9A2A08000097). 
JG is grateful to the Asia Pacific Center for Theoretical Physics for hospitality while this work was under progress.


\begin{thebibliography}{99}




\bibitem{LIGOScientific:2016aoc}
B.~P.~Abbott \textit{et al.} [LIGO Scientific and Virgo],
Phys. Rev. Lett. \textbf{116}, no.6, 061102 (2016)
[arXiv:1602.03837 [gr-qc]].



\bibitem{Bird:2016dcv}
S.~Bird, I.~Cholis, J.~B.~Mu\~noz, Y.~Ali-Ha\"\i{}moud, M.~Kamionkowski, E.~D.~Kovetz, A.~Raccanelli and A.~G.~Riess,
Phys. Rev. Lett. \textbf{116}, no.20, 201301 (2016)
[arXiv:1603.00464 [astro-ph.CO]].



\bibitem{Clesse:2016vqa}
S.~Clesse and J.~Garc\'\i{}a-Bellido,
Phys. Dark Univ. \textbf{15}, 142-147 (2017)
[arXiv:1603.05234 [astro-ph.CO]].



\bibitem{Sasaki:2016jop}
M.~Sasaki, T.~Suyama, T.~Tanaka and S.~Yokoyama,
Phys. Rev. Lett. \textbf{117}, no.6, 061101 (2016)
[erratum: Phys. Rev. Lett. \textbf{121}, no.5, 059901 (2018)]
[arXiv:1603.08338 [astro-ph.CO]].



\bibitem{ZeldovichPBH}
Y.~B. {Zel'dovich} and I.~D. {Novikov}, 
{\em Sov. Astron.}  {\bfseries 10}, 602 (1967).



\bibitem{Hawking:1971ei}
S.~Hawking,
Mon. Not. Roy. Astron. Soc. \textbf{152}, 75 (1971).



\bibitem{Carr:1974nx}
B.~J.~Carr and S.~W.~Hawking,
Mon. Not. Roy. Astron. Soc. \textbf{168}, 399-415 (1974).


\bibitem{Saito:2008jc}
R.~Saito and J.~Yokoyama,
Phys. Rev. Lett. \textbf{102} (2009), 161101
[erratum: Phys. Rev. Lett. \textbf{107} (2011), 069901]
[arXiv:0812.4339 [astro-ph]].



\bibitem{Bugaev:2010bb}
E.~Bugaev and P.~Klimai,
Phys. Rev. D \textbf{83} (2011), 083521
[arXiv:1012.4697 [astro-ph.CO]].



\bibitem{Sasaki:2018dmp}
M.~Sasaki, T.~Suyama, T.~Tanaka and S.~Yokoyama,
Class. Quant. Grav. \textbf{35}, no.6, 063001 (2018)
[arXiv:1801.05235 [astro-ph.CO]].




\bibitem{Harada:2013epa}
T.~Harada, C.~M.~Yoo and K.~Kohri,
Phys. Rev. D \textbf{88}, no.8, 084051 (2013)
[erratum: Phys. Rev. D \textbf{89}, no.2, 029903 (2014)]
[arXiv:1309.4201 [astro-ph.CO]].



\bibitem{Planck:2018vyg}
N.~Aghanim \textit{et al.} [Planck],
Astron. Astrophys. \textbf{641}, A6 (2020)
[erratum: Astron. Astrophys. \textbf{652}, C4 (2021)]
[arXiv:1807.06209 [astro-ph.CO]].



\bibitem{Meszaros:1975ef}
P.~Meszaros,
Astron. Astrophys. \textbf{38}, 5-13 (1975).



\bibitem{Afshordi:2003zb}
N.~Afshordi, P.~McDonald and D.~N.~Spergel,
Astrophys. J. Lett. \textbf{594}, L71-L74 (2003)
[arXiv:astro-ph/0302035 [astro-ph]].



\bibitem{Inman:2019wvr}
D.~Inman and Y.~Ali-Ha\"\i{}moud,
Phys. Rev. D \textbf{100} (2019) no.8, 083528
[arXiv:1907.08129 [astro-ph.CO]].



\bibitem{Carr:2018rid}
B.~Carr and J.~Silk,
Mon. Not. Roy. Astron. Soc. \textbf{478}, no.3, 3756-3775 (2018)
[arXiv:1801.00672 [astro-ph.CO]].



\bibitem{Gong:2017sie}
J.~O.~Gong and N.~Kitajima,
JCAP \textbf{08}, 017 (2017)
[arXiv:1704.04132 [astro-ph.CO]].



\bibitem{Gong:2018sos}
J.~O.~Gong and N.~Kitajima,
JCAP \textbf{11}, 041 (2018)
[arXiv:1803.02745 [astro-ph.CO]].



\bibitem{Mena:2019nhm}
O.~Mena, S.~Palomares-Ruiz, P.~Villanueva-Domingo and S.~J.~Witte,
Phys. Rev. D \textbf{100}, no.4, 043540 (2019)
[arXiv:1906.07735 [astro-ph.CO]].



\bibitem{Planck:2019kim}
Y.~Akrami \textit{et al.} [Planck],
Astron. Astrophys. \textbf{641}, A9 (2020)
[arXiv:1905.05697 [astro-ph.CO]].



\bibitem{Bullock:1996at}
J.~S.~Bullock and J.~R.~Primack,
Phys. Rev. D \textbf{55}, 7423-7439 (1997)
[arXiv:astro-ph/9611106 [astro-ph]].



\bibitem{Byrnes:2012yx}
C.~T.~Byrnes, E.~J.~Copeland and A.~M.~Green,
Phys. Rev. D \textbf{86}, 043512 (2012)
[arXiv:1206.4188 [astro-ph.CO]].



\bibitem{Young:2013oia}
S.~Young and C.~T.~Byrnes,
JCAP \textbf{08} (2013), 052
[arXiv:1307.4995 [astro-ph.CO]].



\bibitem{Young:2014oea}
S.~Young and C.~T.~Byrnes,
Phys. Rev. D \textbf{91} (2015) no.8, 083521
[arXiv:1411.4620 [astro-ph.CO]].



\bibitem{Tada:2015noa}
Y.~Tada and S.~Yokoyama,
Phys. Rev. D \textbf{91} (2015) no.12, 123534
[arXiv:1502.01124 [astro-ph.CO]].



\bibitem{Franciolini:2018vbk}
G.~Franciolini, A.~Kehagias, S.~Matarrese and A.~Riotto,
JCAP \textbf{03} (2018), 016
[arXiv:1801.09415 [astro-ph.CO]].



\bibitem{Kitajima:2021fpq}
N.~Kitajima, Y.~Tada, S.~Yokoyama and C.~M.~Yoo,
JCAP \textbf{10} (2021), 053
[arXiv:2109.00791 [astro-ph.CO]].



\bibitem{Dalal:2007cu}
N.~Dalal, O.~Dore, D.~Huterer and A.~Shirokov,
Phys. Rev. D \textbf{77}, 123514 (2008)
[arXiv:0710.4560 [astro-ph]].



\bibitem{Matarrese:2008nc}
S.~Matarrese and L.~Verde,
Astrophys. J. Lett. \textbf{677}, L77-L80 (2008)
[arXiv:0801.4826 [astro-ph]].



\bibitem{Slosar:2008hx}
A.~Slosar, C.~Hirata, U.~Seljak, S.~Ho and N.~Padmanabhan,
JCAP \textbf{08}, 031 (2008)
[arXiv:0805.3580 [astro-ph]].



\bibitem{Chongchitnan:2010xz}
S.~Chongchitnan and J.~Silk,
Astrophys. J. \textbf{724}, 285-295 (2010)
[arXiv:1007.1230 [astro-ph.CO]].



\bibitem{Gong:2011gx}
J.~O.~Gong and S.~Yokoyama,
Mon. Not. Roy. Astron. Soc. \textbf{417}, 79 (2011)
[arXiv:1106.4404 [astro-ph.CO]].



\bibitem{Politzer:1984nu}
H.~D.~Politzer and M.~B.~Wise,
Astrophys. J. Lett. \textbf{285}, L1-L3 (1984).



\bibitem{Grinstein:1986en}
B.~Grinstein and M.~B.~Wise,
Astrophys. J. \textbf{310}, 19-22 (1986).



\bibitem{Matarrese:1986et}
S.~Matarrese, F.~Lucchin and S.~A.~Bonometto,
Astrophys. J. Lett. \textbf{310}, L21-L26 (1986).



\bibitem{Iliev:2002ms}
I.~T.~Iliev, E.~Scannapieco, H.~Martel and P.~R.~Shapiro,
Mon. Not. Roy. Astron. Soc. \textbf{341} (2003), 81
[arXiv:astro-ph/0209216 [astro-ph]].



\bibitem{Chongchitnan:2012we}
S.~Chongchitnan and J.~Silk,
Mon. Not. Roy. Astron. Soc. \textbf{426} (2012), L21-L25
[arXiv:1205.6799 [astro-ph.CO]].



\bibitem{Sekiguchi:2018kqe}
T.~Sekiguchi, T.~Takahashi, H.~Tashiro and S.~Yokoyama,
JCAP \textbf{02} (2019), 033
[arXiv:1807.02008 [astro-ph.CO]].




\bibitem{Buschmann:2017ams}
M.~Buschmann, J.~Kopp, B.~R.~Safdi and C.~L.~Wu,
Phys. Rev. Lett. \textbf{120} (2018) no.21, 211101
[arXiv:1711.03554 [astro-ph.GA]].



\bibitem{Mishra-Sharma:2020ynk}
S.~Mishra-Sharma, K.~Van Tilburg and N.~Weiner,
Phys. Rev. D \textbf{102} (2020) no.2, 023026
[arXiv:2003.02264 [astro-ph.CO]].



\bibitem{Lee:2020wfn}
V.~S.~H.~Lee, A.~Mitridate, T.~Trickle and K.~M.~Zurek,
JHEP \textbf{06} (2021), 028
[arXiv:2012.09857 [astro-ph.CO]].



\bibitem{Ando:2022tpj}
S.~Ando, N.~Hiroshima and K.~Ishiwata,
Phys. Rev. D \textbf{106} (2022) no.10, 103014
[arXiv:2207.05747 [astro-ph.CO]].






\end{thebibliography}
\end{document}